# Diagnostic results of new-generation dispersive element test samples based on Bragg structures in fused silica

ANTON I. GOROKHOV,[1, *] EVGENY A. PEREVEZENTSEV,[1] MIKHAIL R. VOLKOV,[1] AND IVAN B. MUKHIN[1]

[1]Institute of Applied Physics of the Russian Academy of Sciences, 46 Ul'yanov Street, 603950, Nizhny Novgorod, Russia
*a.gorohov@ipfran.ru

**Abstract:** A comprehensive diagnostics of new-generation dispersive elements based on Bragg structures was completed. The influence of inscription parameters on the properties of samples was studied and a threshold inscription intensity of ~ 6.38 TW/cm$^2$ at which the linear absorption coefficient as well as the phase and polarization stresses sharply increase was found. When inscription is made with an average intensity below the threshold, the linear absorption in dry silica samples remains at a level of $10^{-5}$ cm$^{-1}$, which is approximately two orders of magnitude lower than that of currently available commercial samples. The measured values of the dispersion characteristics using a white light interferometer coincided are in a good agreement with the calculated values provided by the manufacturer. The results of the diagnostics show that the new-generation dispersive elements are promising for compressing kilowatt-level radiation of average power.

## 1. Introduction

In the present-day world, high average power lasers have many applications in both applied and scientific fields. The use of such laser sources allows increasing not only the accuracy of material processing, but also its speed. This is actively employed in various industrial spheres: semiconductor electronics, aviation and automotive industries, modern displays and luxury goods. At the same time, laser systems of this type have a wide range of applicability in scientific research. They are used mainly for pumping in various processes of nonlinear optical transformations, from broadband optical parametric chirped pulse amplification (OPCPA) [1] to the creation of secondary radiation sources of X-ray [2], ultraviolet [3], mid-infrared [4] and terahertz [5] ranges.

However, for most of the above applications, laser systems must have not only a high average power, but also an ultra-short pulse duration. To ensure such a duration, it is necessary to compress the laser radiation, which may be done by means of various dispersive elements. The main problem is the limitation on the use of highly absorbing elements, since at a high repetition rate of laser pulses these elements heat up strongly, resulting in negative thermal effects, such as thermal lens, thermally induced depolarization, self-focusing, and others. Among the most convenient dispersive elements are chirped volume Bragg gratings (CVBGs). Such gratings can provide a high dispersion at small sizes, and are also easy to align compared, for example, to the widely used diffraction gratings. Today, the main and leading technology for the production of CVBGs is based on exposure to ultraviolet light and thermal post-processing of photo-thermo-refractive glass [6]. The main disadvantage of such gratings is the high absorption coefficient. In this regard, the record advance for commercial CVBGs made of photo-thermo-refractive glass is the compression of radiation with an average power of about 150 W [7]. However, achieving such values is accompanied with a number of serious problems, the main of which is active thermal stabilization of CVBGs significantly complicating the system. A further increase in the average power can lead to substantial distortions of the laser pulse and of dispersion in the CVBG, and even to the destruction of the grating.

At the same time, there are currently quite a lot of commercial and prospective laser systems under way with an average power exceeding the maximum allowable one for compression by available dispersive elements [8-11]. Therefore, the creation of dispersive elements with low absorption is a highly relevant task. In this regard, methods of inscribing that allow overcoming the material barrier are currently being actively developed. They will enable fabricating gratings from materials other than photorefractive ones, which have a significantly lower absorption. An advanced inscription method is to use femtosecond laser pulses and a phase mask [12].

This method was employed by a group of scientists from the Institute of Applied Physics (IAP) at the Friedrich Schiller University Jena who produced a set of dispersive elements based on Bragg structures made of fused silica with various inscription parameters, and transferred them to the Institute of Applied Physics of the Russian Academy of Sciences (IAP RAS) for studying their laser and optical characteristics. The set included chirped volume Bragg gratings (Fig. 1, b) and volume Bragg gratings (VBG) with a constant period, which were used as a transmission diffraction grating (Fig. 1, a).

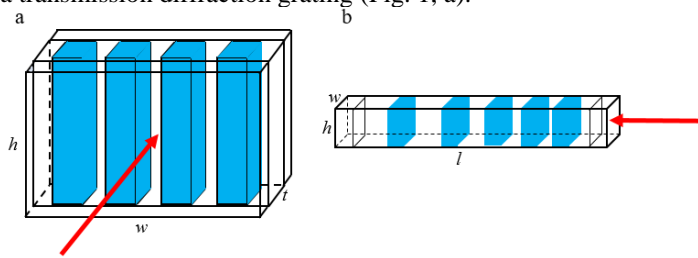

Fig. 1. Schematic representation of radiation injection into VBG used as a transmission diffraction grating and CVBG. Grooves are shown in bright blue.

The material was modified using a Ti:sapphire laser at a maximum repetition rate of 10 kHz and an average output power of 7 W by ultrashort pulses emitted with a minimum duration of 100 fs at a central wavelength of 800 nm. Extensive diagnostics of the first batch of fused silica CVBGs was made and described in the paper [13]. The results confirmed that they are promising for use in systems with high average power, but some parameters have not been fully investigated yet and require additional measurements. For further research, our colleagues from the IAP at the Friedrich Schiller University Jena transferred to us another batch of gratings inscribed in dry silica with lower absorption. These samples are the first comprehensively diagnosed volumetric dispersive elements inscribed in fused silica using femtosecond laser pulses with a phase mask. In this paper we describe the studies of the laser and optical characteristics of the above samples.

The influence of the parameters and schemes of material modification on grating characteristics are investigated. The obtained results and the revealed regularities are described in the first section of this work. The next two sections are devoted to the improvement of some methods of measuring grating characteristics: the expansion of the phase distortion measurement technique is described in the second section and the dispersion of the CVBG is studied in the final section using direct measurements of group delay from the wavelength with a white light interferometer. Also, the measurements of the diffraction efficiency for the VBG used as a transmittance diffraction grating are described.

## 2. Influence of parameters and schemes of material modification on grating characteristics

When dispersive elements are manufactured by the phase mask method, various radiation inscription modes with different characteristics can be used, which ultimately affect the output parameters of the gratings. Since the use of such elements in real systems requires scaling their size, the inscription technique needs improvement. To determine the most optimal parameters and inscription modes, a number of experiments were conducted aimed at revealing the dependence of grating characteristics on the parameters and modes of material modification. To improve measurement accuracy and reduce thermal effects, the samples inscribed in dry fused silica with a low intrinsic linear absorption coefficient were used. Gratings inscribed at different laser radiation energies, as well as with different inscription speeds, pulse repetition rates, and scanning steps were used for comparison (Table 1). The gratings in all samples were inscribed for operating at a central wavelength of 1030 nm. The VBGs have a characteristic cross-sectional size (w×h) of 5-10 mm and a thickness (t) of several millimeters (Fig. 1, a). The CVBGs have a cross section (w×h) with sides of ~2 mm and a length (l) of 30-40 mm (Fig. 1, b).

|  | A1 | T2 | T1 | A2 |
|---|---|---|---|---|
| Inscription energy $E_p$, μJ | **500** | **550** | **600** | **650** |
| Rep. rate, kHz | 10 | 10 | 10 | 5 |
| Scanning speed, mm/s | 0.2 | 1 | 1, 2, 3 (top to bottom) | 0.05 |
| Number of scans | 3 | 400 | 600 | 1 |
| Scan separation, mm | 3 | 0.05 | 0.05 | - |
| Depth, mm | 2 | 2 | 1.5 | 2 |

Table 1. Inscription parameters.

First of all, we measured the linear absorption coefficient using a unique calorimetric method [14]. The measurement results are shown in Fig. 2: the circles correspond to the measurements taken at a specific point, but in some areas indicated by rectangles the values were variable, so average absorption values are given for them. The studies revealed an interesting correlation between the grating parameters and the energy of the inscribing pulses $E_p$. The sample obtained at $E_p$ = 500 μJ has the same absorption coefficient in the modified and unmodified areas ~$10^{-5}$ 1/cm, which corresponds to the intrinsic absorption of the material (Fig. 2, a). At $E_p$ = 550 μJ, the absorption coefficient increases in some areas of the inscribed region, but on the average, it remains at the $10^{-5}$ 1/cm level (Fig. 2, b). The next sample includes three gratings inscribed at $E_p$ = 600 μJ, but with different scanning speeds (Fig. 2, c). It is seen that all gratings have the same absorption of ~$10^{-4}$ 1/cm. This indicates that the inscription speed does not affect the absorption. The gratings inscribed with $E_p$ = 650 μJ also have absorption at the $10^{-4}$ 1/cm level (Fig. 2, d). Thus, a significant increase in the absorption coefficient can be observed at the inscription energy of ~550 μJ.

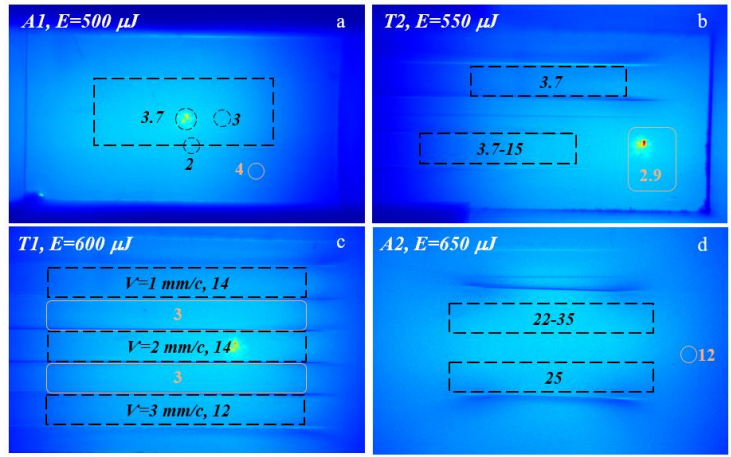

Fig. 2. Results of measuring linear absorption coefficient for samples of dry silica A1(a), T2(b), T1(c), and A2(d). Black dashes delimit grating areas. Orange lines indicate unmodified areas.

Next, we studied the influence of the parameters and schemes of material modification on the polarization distortions of radiation inside the sample. The measurements were made using the crossed-polarizers technique with image transfer from the plane of the samples to the camera matrix. The local depolarization distributions for samples inscribed with pulses of different energies are shown in Fig. 3. It is clearly seen that with an increase in energy from 500 to 550 µJ, the depolarization sharply increases by almost 2 orders of magnitude. A further increase in the inscription energy increases the depolarization only slightly. This is due to the fact that at an energy of 550 µJ the depolarization almost reaches its theoretical threshold of 50% and cannot increase further.

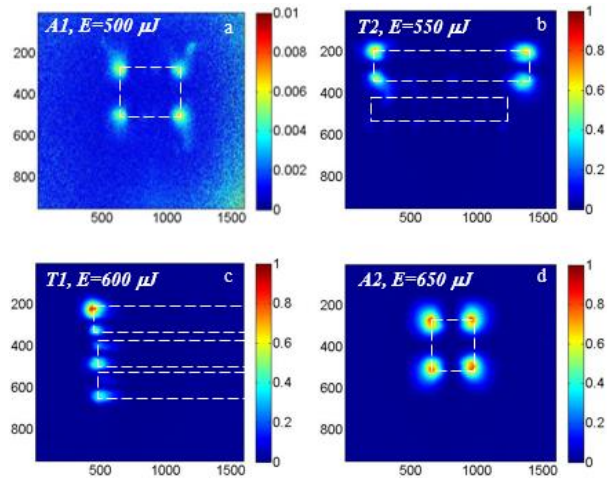

Fig. 3. Local depolarization for samples of dry quartz A1(a), T2(b), T1(c), and A2(d). Dashes delimit grating area.

Based on the obtained results, we concluded that the inscription energy of 550 µJ equivalent to the intensity of ~ 6.38 TW/cm$^2$ seems to be the threshold, after which the absorption in the modified areas starts to increase and the depolarization also sharply increases. It is important to note that, according to the results of the studies, no dependence of linear absorption, polarization distortions, or other grating characteristics on other inscription parameters was revealed. The absorption values, even in the regions inscribed at high energy, were ~10$^{-4}$ 1/cm, which is approximately an order of magnitude lower than that of commercial CVBGs made of photo-thermo-refractive glass and is an indicator of the prospects for using the new type of gratings in systems with high average power. At the same time, in most samples polarization distortions were high. However, as was found in the previous work, this problem can be solved by cutting the grating out of the area of unmodified silica, which allows reducing the integral depolarization by more than 2 orders of magnitude [13].

## 3. Studying phase distortions

The studies described in our earlier work revealed mechanical stresses arising in the gratings in the course of their inscription, which led to polarization and phase distortions. We made a comprehensive investigation of the polarization distortions, whereas the phase distortions were only recorded using a qualitative method. Quantitative results for new VBG samples made of dry silica have been obtained with phase-shifting interferometry [15]. Initially, one of the grating planes had a more pronounced concavity (Fig.4, a) relative to the other (Fig.4, b). This difference is caused by the fact that the grating inside the volume of unmodified material is pressed closer to one of the planes, hence, the mechanical stresses on this plane are stronger, which leads to a stronger surface curvature.

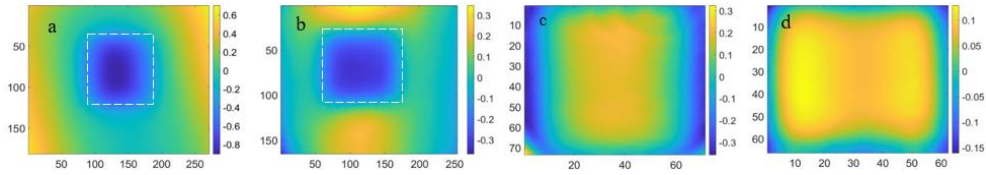

Fig.4. VBG surface profiles (delimited by dashes) (a,b) prior to cutting to 4×4 mm size (c,d); μm color scale.

As shown in the earlier studies, mechanical stresses resulting in phase distortions may be eliminated by cutting the grating out of the bulk of unmodified silica. After cutting the grating to a size of 4×4 mm, the larger central part of both planes, which will be the operating region, can be considered to be almost flat (Fig. 4 c, d). It is important to note that these distortions may be partially related to the curvature of the surface of the original sample in which the inscription is performed.

## 4. Studying dispersive characteristics

The final step towards completing the comprehensive CVBG diagnostics was studying its dispersive characteristics. The measurements using the autocorrelation technique proved to be inadequate for obtaining reliable results as they did not allow constructing an approximating function of sufficient accuracy. To study this problem in more detail, direct measurements of the dependence of the group delay on the wavelength were carried out using a white light interferometer [16] (Fig. 5, a).

The source radiation is first incident on the beam splitter, after which half of the radiation propagates to the grating, and the other half is incident on the flat mirror located on the translation stage. Then both beams return back to the beam splitter, and the resulting signal is fed into a ~15-pm high-resolution spectrometer (LightMachinery HN-9354). By moving the mirror on the translation stage, it is possible to change the length of one of the interferometer arms. When the mirror position is changed, the depth of signal reflection inside the CVBG changes too. As a result, the wavelength at which interference occurs and the corresponding central peak of the distribution observed on the spectrometer change as well (Fig. 5, b). The theoretical distribution of intensity $I$, the positions of the minima and maxima of which coincide with the experimental one, is constructed using the expressions (1, 2)

$$I(\lambda) = I_0(\lambda)\left(1 + \cos\pi\left(\frac{\lambda-\lambda_0}{\Omega}\right)^2\right) \qquad (1)$$

$$\Omega = \sqrt{\frac{4\pi^2 c}{D\lambda_0^2}},\qquad(2)$$

where $I_0$ is the maximum intensity, $\lambda_0$ is the wavelength at which reflection occurs, $D$ is the grating dispersion, and $c$ is the speed of light in vacuum.

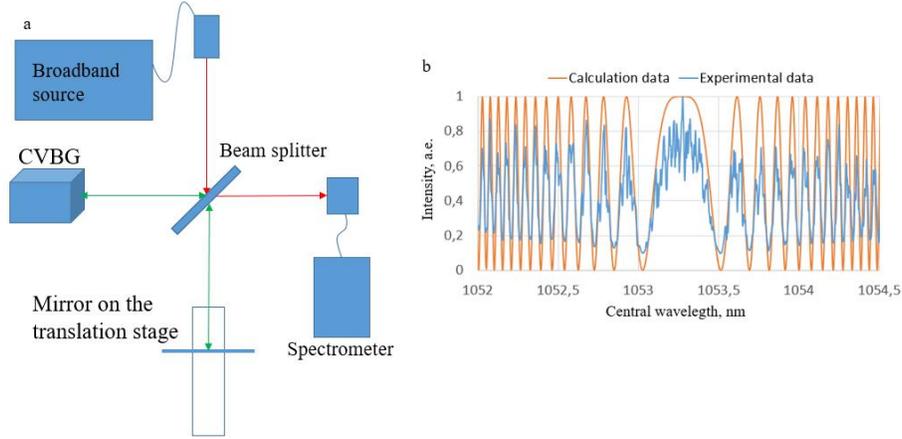

Fig. 5. Schematic diagram of measuring the dependence of group delay in CVBG on wavelength with white light interferometer (a). Signal intensity at the interferometer output versus wavelength (b).

By studying the dependence of the wavelength corresponding to the interference maximum on the mirror location, we can obtain the dependence of the delay on the wavelength. The central wavelength was found as the average value for the minima of the same order, but of different sign in Fig. 5, b. It was ascertained experimentally that for higher accuracy it is necessary to additionally average several values of the central wavelength obtained for different orders of minima.

To verify the accuracy of this method we first measured dispersion of the OptiGrate gratings, since they are commercial gratings with known technical data. Two samples with spectrum widths of 8 nm (Fig. 6, a) and 2.24 nm (Fig.6, b) were measured.

The dispersion of the 8-nm grating was 62±1 ps/nm, which is very close to the value of 60±5 ps/nm declared by the manufacturer. The dispersion of the 2.24-nm grating was 207±1 ps/nm, which also matched to a high accuracy the dispersion declared by the manufacturer 200±5 ps/nm. These results show that the white light interferometer was assembled and correctly aligned, which makes it a good tool for accurate measurements of the dispersion characteristics of new-generation gratings.

Next, we measured two dry silica gratings inscribed by femtosecond laser pulses. The T1 grating had a dispersion of 0.3262 nm/mm (Fig.6, c), which is actually the same as the value of 0.32 nm/mm obtained from the manufacturer. The T2 grating had a dispersion of 0.2738 nm/mm (Fig. 6, d), which differs from the manufacturer's data by approximately 15%. The reasons for this difference will be discussed with the manufacturer.

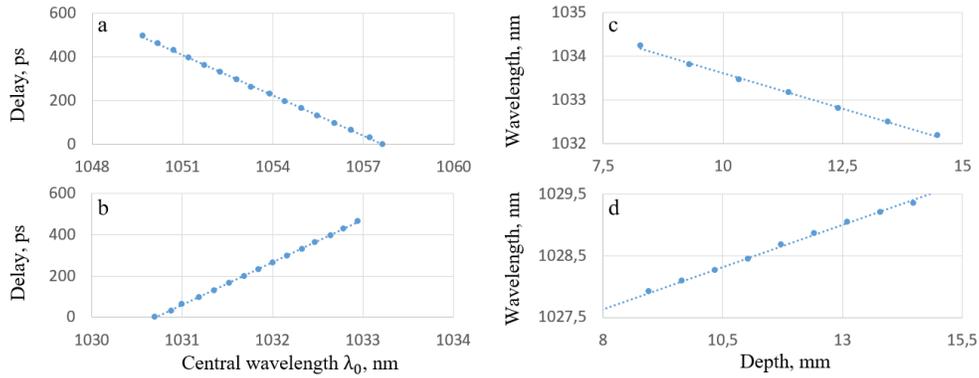

Fig.6. Group delay versus central wavelength for OptiGrate gratings with 8-nm (a) and 2.24-nm (b) spectrum width measured by the interference technique. Wavelength of reflected laser pulse versus reflection depth inside T1(c) and T2 (d) gratings.

As a result of the made diagnostics of the new generation CVBGs, a low absorption coefficient, small distortions and a good agreement of the dispersion characteristics with the manufacturer's data have been revealed. The overall results described above, as well as those obtained earlier [13], show the prospects for using such gratings in systems with high average power.

By way of additional studies, we measured the diffraction efficiency η for the VBGs used as transmission diffraction gratings. The measurements were performed by the scheme shown in Fig. 7. The scheme was adjusted so that the radiation incidence angle $\alpha$ was close to the diffraction angle $\beta$. Therefore, the expression (3) could be simplified to the form (4) and the diffraction efficiency was calculated by the formula (5):

$$d(sin\alpha + sin\beta) = m\lambda \qquad (3)$$

$$2d sin\alpha = m\lambda, \qquad (4)$$

where the grating period $d$ = 1.7765 μm, the central wavelength $\lambda$ = 1073 nm, $\alpha$ is the angle of incidence on the grating, $\beta$ is the diffraction angle, $m$ is the order of diffraction, and

$$\eta = \frac{P_{dif}}{P_{trans} + P_{dif}}, \qquad (5)$$

where $P_{dif}$ is the power of the diffracted radiation, and $P_{trans}$ is the power of the transmitted radiation.

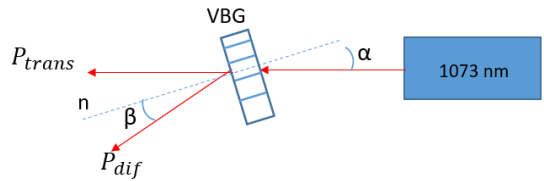

Fig. 7. Scheme of diffraction efficiency measurements.

The diffraction efficiency of different grating orders inscribed by pulses with an energy of 650 μJ is presented in Table 2. The results demonstrate that in the first diffraction order, the grating has a fairly high diffraction efficiency of 82%, but at higher orders it drops significantly. It is important to note that the calculations took into account reflections from the grating faces.

| m | η |
|---|---|
| 1 | 0.82 |
| 2 | 0.55 |
| 3 | 0.32 |

Table 2. Dependence of diffraction efficiency on the order of diffraction at the angle of incidence on the grating close to the angle of diffraction.

A compression scheme of high-average-power laser radiation was implemented experimentally. The scheme consists of a grating and two retroreflective mirrors providing four passes of radiation through the grating. A source with a spectrum width close to the width of Yb:YAG lasers was used for measurements, since these lasers are used in most present-day lasers with high average power. The input radiation had a symmetric spatial intensity distribution (Fig. 8, b). A sufficiently high beam quality was maintained at the output after four passes (Fig. 8, c), which is indicative of small spatial distortions and, consequently, of the prospects for continuing work in this direction. The average diffraction efficiency per pass was approximately 0.58, which, in our opinion, may be explained by diffraction losses due to the small size of the test sample. The radiation spectrum at the output almost completely coincided with the spectrum at the input (Fig. 8, d).

The studies conducted on the basis of a four-pass scheme of laser beam compression confirmed that such gratings are promising for use in systems with high average power.

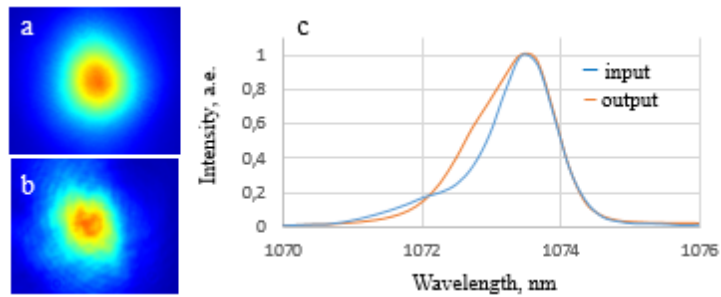

Fig. 8. Beam profiles at the input (a) and output (b). Spectra of input (blue) and output (orange) radiation (c).

## 5. Conclusion

The set of diagnostic studies of new-generation dispersive elements based on Bragg structures has been finalized. The main competitive advantage of such elements is low absorption at the level of $10^{-4} - 10^{-5}$ 1/cm, which is at least an order of magnitude lower than that of currently available commercial samples. To upgrade the technology of inscribing new CVBGs, the influence of the parameters and schemes of material modification on the properties of the obtained samples has been studied. For the used inscription configuration, a threshold inscription intensity ~ 6.38 TW/cm$^2$ at which the linear absorption coefficient as well as the phase and polarization stresses sharply increase was found. However, even in this case, the absorption does not rise above $10^{-4}$. This allows us to conclude that the new CVBGs are highly promising for compression of kilowatt-level radiation of average power. With the use of the high-precision technique of phase-shifting interferometry, quantitative estimates of phase distortions arising in the course of grating inscription have been obtained. The dispersion characteristics have been measured with a white light interferometer. The obtained results have a good agreement with the calculated ones. We have also measured the diffraction efficiency of the VBGs used as transmission diffraction gratings. Based on such gratings, a four-pass compression scheme with low distortions has been assembled. The average diffraction efficiency per pass of 0.58 has been obtained. The results of the conducted diagnostics clearly show the potential of the new generation of dispersive elements. However, the technology of the inscription still requires updating, primarily aimed to improve aperture homogeneity.

**Funding.** Studying dispersion characteristics was supported by Russian Science Foundation No. 24-12-00461, https://rscf.ru/project/24-12-00461/. Investigation of influence of parameters and modes of material modification on gratings linear absorption, polarization and phase distortions was supported by the State Research Task for the Institute of Applied Physics Russia Academy of Sciences (Project No. FFUF-2024-0043).

**Acknowledgment.** We thank our colleagues from the Institute of Applied Physics at the Friedrich Schiller University Jena for providing samples with a description of their characteristics and inscription parameters.